# Design of robust $H_\infty$ fuzzy output feedback controller for affine nonlinear systems: Fuzzy Lyapunov function approach

Leila Rajabpour*, Mokhtar Shasadeghi**
and Alireza Barzegar***

Department of Electronic Engineering

*University of Technology Malaysia, Malaysia

**Shiraz University of Technology, Shiraz, Iran

*** Nanyang Technological University, Singapore

E-mail: l.rajabpoor@gmail.com

E-mail: shasadeghi@sutech.ac.ir

E-mail: alireza001@e.ntu.edu.sg

**Abstract:** In this paper, we propose a new systematic approach based on non-quadratic Lyapunov function and technique of introducing slack matrices, for a class of affine nonlinear systems with disturbance. To achieve the goal, first, the affine nonlinear system is represented via Takagi–Sugeno (T–S) fuzzy bilinear model. Subsequently, the robust $H_\infty$ controller is designed based on parallel distributed compensation (PDC) scheme. Then, the stability conditions are derived in terms of linear matrix inequalities (LMIs) by utilizing Lyapunov function. Moreover, some slack matrices are proposed to reduce the conservativeness of the LMI stability conditions. Finally, for illustrating the merits and verifying the effectiveness of the proposed approach, the application of an isothermal continuous stirred tank reactor (CSTR) for Van de Vusse reactor is discussed in details.

**Biographical notes:** Leila Rajabpour received her B.S. degree in Electronic Engineering from Shiraz University of Technology, Shiraz, Iran in 2010. She is a current Master student at University of Technology Malaysia (UTM). Her research interests include intelligent systems, fuzzy control, electrical system control and robust control.

Mokhtar Shasadeghi received his B.S. degree in Electronics Engineering from Shiraz University, Shiraz in 1996, and his M.Sc. and Ph.D. degrees from Tarbiat Modares University, in 2001 and 2007, respectively, all in Iran. His research interests include robust control, adaptive control, fuzzy control, time delay systems, optimization, LMI, and neural networks.

Alireza Barzegar received his B.S. degree in Electrical Engineering from Shiraz University of Technology, Shiraz, Iran in 2010 and M.Sc. degrees in Electrical

Engineering, Control and System Engineering, from Khajeh Nasir Technological University, Tehran, Iran in 2012. Currently, he is a Ph.D. candidate at Nanyang Technological University (NTU), Singapore. His research interests include relaxation and optimization of optimal power flow of electrical systems, intelligent systems and fuzzy control.

# 1 Introduction

Stability problem and control of the nonlinear systems has always been an important issue to the control scientists. Lots of these problems cannot be solved only by using linear techniques and the need to more advanced technologies leads to the formation of developed nonlinear control methods such as methods proposed by Khalil (2002). Recently, the well-known Takagi–Sugeno (T–S) fuzzy model, which is a simple and effective tool in control of complex nonlinear systems has attracted a lot of attention, as pointed out by Sha Sadeghi et al. (2016) and Wei et al. (2016). Additionally, it may provide an exact representation of the nonlinear system (Tanaka and Wang 2001). The fuzzy control via PDC through the Lyapunov theorem is the leading approach in stability analysis and controller design for T–S fuzzy systems (Li et al. 2008). By using the PDC control approach we can investigate the stability problem in the form of LMIs. It is shown that the conservativeness of LMI-based stability conditions is a big problem when deriving the stability condition based on quadratic Lyapunov function (QLF).

Non-quadratic (Fuzzy) Lyapunov function is considered as a solution to solve the conservativeness problem of the QLF-based LMIs. Some of the studies based on NQLF approach for T–S fuzzy systems have been addressed in Guerra et al. (2012) and Sha Sadeghi and Vafamand (2014). The NQLF is the fuzzy combination of a number of QLFs, which leads to appearance of the time derivative of membership functions (MFs) and their upper bounds in LMIs. Due to the direct effect of the upper bounds of MFs on the speed of answer, they could not be chosen by trial and error and should be determined in an optimal manner. To overcome this problem, the upper bounds of the MFs are considered as an LMI variable as in Vafamand and Sha Sadeghi (2015) and in order to convert the upper bounds of time derivative of MFs as a decision variable a generalized-eigenvalue-problem (GEVP) is used.

In addition to the fuzzy control, it is known that bilinear systems proposed by Elliott (1999), as an important class of nonlinear systems, give a better approximation of the nonlinear systems than the linear ones, so, they have been successfully applied to many real-world systems, including many physical systems and chemical processes in engineering fields (Rahmani et al., 2017; Lee et al., 2017). As pointed out by Hamdy and Hamdan (2015), a nonlinear system can be modeled as bilinear system as below:

$$\dot{x}(t) = f(x(t), u(t)) = Ax(t) + Bu(t) + Nx(t)u(t) \tag{1}$$

where $x(t) \in R^{n \times 1}$ is the state vector, $u(t) \in R$ is the control input, $A \in R^{n \times n}, B \in R^{n \times 1}$ and $N \in R^{n \times n}$ are known matrices. As can be seen, a bilinear system involves products of state and control which means that they are linear in state and linear in control but not jointly linear in control and state. In fact, a bilinear system exists between nonlinear and linear systems.

Because of the advantages of T–S fuzzy model-based bilinear system, it becomes one of the recent interesting research platforms in model-based fuzzy control (Yang et al., 2016; Chang and Hsu, 2016). The robust $H_\infty$ problem for continuous-time fuzzy bilinear system (FBS) was first proposed by Li and Tsai (2007). Since then, numerous studies on the stability analysis and control of FBS have been done (Hamdy and Hamdan, 2014; Yoneyama, 2017). Very few of literatures considered the output feedback control for discrete FBS (Yu and Jo, 2016) and continuous-time FBS (Hamdy and Hamdan, 2015; Hamdy et al., 2014), while in many practical cases, the states are not available for controller implementation, therefore, in such cases, output feedback controller is necessary. It is also concluded that in some cases the implementation of output feedback controller is cheaper and simpler in construction and maintenance (Hamdy and Hamdan, 2015).

In this paper, a novel approach for stabilizing the continuous-time FBS was conducted with benefits of output feedback control and NQLF approach. To the best of our knowledge, no previous study has derived the stability condition of the continuous-time bilinear system via output feedback based on the NQLF approach.

In addition, in this paper by introducing a new slack matrix we will obtain more relaxed stability conditions. The structure of the slack matrix is chosen in a way that facilitates the proof procedure of the proposed approach.

In brief, the main focus of this study will be on: designing a robust fuzzy output feedback controller for the continuous-time FBS with disturbance, deriving the LMIs conditions for the stability analysis of the FBS based on fuzzy Lyapunov function. Moreover, some null terms are proposed to introduce a slack matrix to drive new stability conditions and finally, utilizing a GEVP in order to convert the upper bounds of time derivative of membership functions as a decision variable.

The remainder of the paper is organized as follows. The problem formulation and the robust output feedback controller via PDC approach by considering the idea of fuzzy Lyapunov function are stated in Section 2. Stabilization conditions are derived in terms of LMIs in Section 3. In Section 4 Simulations results are provided to illustrate the effectiveness of our proposed approach. Finally, conclusions are given in Section 5.

## 2   Problem Formulation

Consider a class of nonlinear system with affine input variables as follows:

$$\dot{x}(t) = f(x(t), u(t)) = F(x(t)) + G(x(t))u(t) + Nx(t)u(t) + Ew(t) \qquad (2)$$

where $x(t) \in R^{n \times 1}$ is state vector, $u(t) \in R$ is the control input, $w(t) \in R^{m \times 1}$ is the disturbance input, $F(x(t)) \in R^{n \times 1}, G(x(t)) \in R^{n \times 1}, N \in R^{n \times n}$ and $E \in R^{n \times m}$.

Then similar to Khalil (2002), by approximating the behavior of the nonlinear system with disturbance (2) in neighborhood of the desired operating point $x_d$, the T–S fuzzy model with disturbance is derived as follows:

Plant rule $i$: IF $s_1(t)$ is $M_{1i}$ and ... ... and $s_v(t)$ is $M_{vi}$

$$\text{THEN } \begin{cases} \dot{x}(t) = A_i x(t) + B_i u(t) + N_i x(t) u(t) + E_i w(t) \\ z(t) = C_{1i} x(t) + D_i w(t) \\ y(t) = C_{2i} x(t) \end{cases} \qquad (3)$$

where $r$ is the number of rules, $M_{ji}$ $(i = 1, 2, ..., r, j = 1, 2, ..., v)$ is the fuzzy set and $s_1(t), s_2(t), ..., s_v(t)$ are the known premise variables. Each bilinear consequent equation represented by $A_i x(t) + B_i u(t) + N_i x(t) u(t) + E_i w(t)$ is known as a subsystem. $x(t) \in R^{n \times 1}$ is the state vector, $u(t) \in R$ is the control input, $z(t) \in R$ is the controlled output, $w(t) \in R^{m \times 1}$ is the disturbance input and $y(t) \in R$ is the measured output. The matrices $A_i \in R^{n \times n}, B_i \in R^{n \times 1}, N_i \in R^{n \times n}, E_i \in R^{n \times m}, C_{1i} \in R^{1 \times n}, D_i \in R^{1 \times m}$ and $C_{2i} \in R^{1 \times n}$ are known with appropriate dimensions, for $i = 1, 2, ..., r$.

By using singleton fuzzifier, product inference and center average defuzzifier, one can get the following overall fuzzy bilinear model:

$$\dot{x}(t) = \frac{\sum_{i=1}^{r} \alpha_i(s(t)) (A_i x(t) + B_i u(t) + N_i x(t) u(t) + E_i w(t))}{\sum_{i=1}^{r} \alpha_i(s(t))}$$

$$= \sum_{i=1}^{r} h_i(s(t))\big(A_i x(t) + B_i u(t) + N_i x(t) u(t) + E_i w(t)\big)$$

$$z(t) = \sum_{i=1}^{r} h_i(s(t))\big(C_{1i} x(t) + D_i w(t)\big) \quad (4)$$

$$y(t) = \sum_{i=1}^{r} h_i(s(t)) C_{2i} x(t)$$

where $\alpha_i(s(t)) = \prod_{j=1}^{v} M_{ji}(s_j(t))$ and $h_i(s(t)) = \frac{\alpha_i(s(t))}{\sum_{i=1}^{r} \alpha_i(s(t))}$ for all $t$. $M_{ji}(s_j(t))$ is the membership grade of $s_j(t)$ in $M_{ji}$ and $s(t) = [s_1(t), \dots, s_v(t)]^T \in R^{v \times 1}$. Since $\alpha_i(s(t)) \geq 0$, $\sum_{i=1}^{r} \alpha_i(s(t)) > 0$, $i = 1, 2, \dots, r$, then we have $h_i(s(t)) \geq 0$ and $\sum_{i=1}^{r} h_i(s(t)) = 1$ for all $t$.

The overall robust fuzzy output feedback controller for stabilizing the T–S fuzzy bilinear model with disturbance (4) via PDC technique can be formulated as follows:

$$u(t) = \sum_{i=1}^{r} h_i(s(t)) \frac{\beta k_i y(t)}{\sqrt{1 + (k_i y(t))^2}}$$

$$= \sum_{i=1}^{r} h_i(s(t)) \beta k_i y(t) \cos\theta_i$$

$$= \sum_{i=1}^{r} h_i(s(t)) \beta \sin\theta_i \quad (5)$$

where

$$\sin\theta_i = \frac{k_i y(t)}{\sqrt{1 + (k_i y(t))^2}}, \quad \theta_i \in \left[-\frac{\pi}{2}, \frac{\pi}{2}\right]$$

$$\cos\theta_i = \frac{1}{\sqrt{1 + (k_i y(t))^2}}, \quad \theta_i \in \left[-\frac{\pi}{2}, \frac{\pi}{2}\right]$$

$k_i$ is a scalar to be determined and $\beta > 0$ is a scalar can be arbitrary designed.

By substituting (5) into (4), the following closed loop system is obtained:

$$\dot{x}(t) = \sum_{i=1}^{r}\sum_{j=1}^{r}\sum_{l=1}^{r} h_i(s(t)) h_j(s(t)) h_l(s(t)) \big(A_i x(t) + B_i \beta k_j C_{2l} x(t) \cos\theta_j + N_i x(t) \rho \sin\theta_j + E_i w(t)\big) \quad (6)$$

and by rewriting (6), we have:

$$\dot{x}(t) = \sum_{i=1}^{r}\sum_{j=1}^{r}\sum_{l=1}^{r} h_i(s(t)) h_j(s(t)) h_l(s(t)) \big((A_i + \beta B_i k_j C_{2l} \cos\theta_j + \rho N_i \sin\theta_j) x(t) + E_i w(t)\big) \quad (7)$$

The robust $H_\infty$ fuzzy output feedback control problem can be formulated as follows:
1) The closed-loop system (7) is asymptotically stable when $w(t) = 0$.
2) Given the fuzzy system (7) and a scalar $\gamma > 0$, under zero initial condition $x(0) = 0$, the controlled output $z(t)$ satisfies

$$\|z(t)\|_2 < \gamma \|w(t)\|_2 \tag{8}$$

for any nonzero $w(t) \epsilon L_2[0, \infty]$.

Next, to show that (7) satisfies (8), we introduce

$$J = \int_0^\infty [z(t)^T z(t) - \gamma^2 w(t)^T w(t)]\, dt \tag{9}$$

## 3  Main Results

Our purpose is deriving a non-quadratic stabilization condition for the T–S FBS with disturbance, the approach is based on the idea of NQLF. By considering the maximum upper bounds of MFs as an LMI variable and introducing a new slack matrix, more relaxed stability conditions are obtained. Hence, our approach leads to more applicability in control design.

The following slack matrix is proposed to obtain more relaxed stability conditions and also it is a great help in conversion of the stability problem into an LMI form. From $\sum_{i=1}^r h_i(s(t)) = 1$ and then $\sum_{i=1}^r \dot{h}_i(s(t)) = 0$, it is concluded that there exists slack matrix $M$ such that

$$\sum_{i=1}^r \dot{h}_i(s(t)) \left\{ x^T(t) \left( \frac{M}{r} - \sum_{k=1}^r \frac{P_k}{r} \right) x(t) \right\} = 0 \tag{10}$$

where $P_k, k = 1, \dots, r$ is a positive definite matrix and $M$ is a matrix with appropriate dimensions. The structure of this slack matrix is chosen in an innovative way so that it helps simplifying the proof process.

The following lemma is well-known and will be very useful in the proof of our main results.

**Lemma1**. For any two matrices $X$ and $Y$ with appropriate dimensions, and $\varepsilon > 0$, we have

$$X^T Y + Y^T X \le \varepsilon X^T X + \varepsilon^{-1} Y^T Y. \tag{11}$$

Our new LMI fuzzy Lyapunov function-based approach for solving the robust $H_\infty$ fuzzy output feedback control of fuzzy bilinear system with disturbance (7) is described in following.

**Theorem.** Consider the time derivative of MFs has an upper bound such that:

$$|\dot{h}_\rho| < \phi_\rho \le \Phi \qquad \rho = 1, \dots, r \tag{12}$$

The closed-loop T–S fuzzy bilinear model with disturbance (7) is stable via robust fuzzy output feedback controller (5) if there exist positive definite matrices $P_i = P_i^T$, $i = 1, 2, \dots, r$, a scalar $\rho > 0$, some scalars $\varepsilon_{ijl} > 0$ and control gains $k_j$ such that the following LMIs are satisfied:

$$P_\rho + \frac{M}{r} - \sum_{k=1}^r \frac{P_k}{r} > 0, \quad \rho, k = 1, 2, \dots, r \tag{13}$$

$$P_e > 0, \qquad e = 1, 2, \dots, r \tag{14}$$

$$\begin{bmatrix} (A_i^T P_e + *) + \varepsilon_{ijl} N_i^T N_i + C_{1i}^T C_{1i} + \Phi M & C_{1i}^T D_i + P_e^T E_i & (B_i k_j C_{2l})^T & P_e \\ * & D_i^T D_i - \gamma^2 I & 0 & 0 \\ * & * & -\varepsilon_{ijl}^{-1} I & 0 \\ * & * & * & -\varepsilon_{ijl} \rho^{-2} I \end{bmatrix} < 0 \tag{15}$$

where $i, j, l, e = 1, \dots, r$ and * denotes the transposed elements in the symmetric positions.

**Proof:** To obtain stability conditions, non-quadratic Lyapunov function and robust output feedback controller is utilized. By considering the non-quadratic Lyapunov function (16) and slack matrix (10), one has:

$$V(x(t)) = \sum_{i=1}^{r} h_i(s(t)) x(t)^T P_i \, x(t) \tag{16}$$

then,

$$\begin{aligned}
\dot{V} &= \left\{ \dot{x}^T \left( \sum_{i=1}^{r} h_i P_i \right) x + * \right\} + x^T \left( \sum_{\rho=1}^{r} \dot{h}_\rho P_\rho \right) x \\
&\quad + \sum_{\rho=1}^{r} \dot{h}_\rho \, x^T \left\{ \left( \frac{M}{r} - \sum_{k=1}^{r} \frac{P_k}{r} \right) \right\} x \\
&= \sum_{i=1}^{r} h_i \left[ \{ \dot{x}^T P_i x + * \} + x^T \sum_{\rho=1}^{r} \dot{h}_\rho \left\{ P_\rho + \frac{M}{r} - \sum_{k=1}^{r} \frac{P_k}{r} \right\} x \right]
\end{aligned} \tag{17}$$

Suppose (12) holds. If

$$P_\rho + \frac{M}{r} - \sum_{k=1}^{r} \frac{P_k}{r} \geq 0 \tag{18}$$

also holds, then we have:

$$\begin{aligned}
\sum_{\rho=1}^{r} \dot{h}_\rho \left\{ P_\rho + \frac{M}{r} - \sum_{k=1}^{r} \frac{P_k}{r} \right\} &\leq \sum_{\rho=1}^{r} \Phi \left\{ P_\rho + \frac{M}{r} - \sum_{k=1}^{r} \frac{P_k}{r} \right\} \\
&= \Phi \sum_{\rho=1}^{r} \left\{ P_\rho + \frac{M}{r} - \sum_{k=1}^{r} \frac{P_k}{r} \right\} = \Phi \left\{ \sum_{\rho=1}^{r} P_\rho + \sum_{\rho=1}^{r} \frac{M}{r} - \sum_{\rho=1}^{r} \sum_{k=1}^{r} \frac{P_k}{r} \right\} \\
&= \Phi \left\{ \sum_{\rho=1}^{r} P_\rho + M - \sum_{k=1}^{r} P_k \right\} = \Phi M
\end{aligned} \tag{19}$$

By substituting the above result in (17) and considering the closed loop system (7), (17) is continued as:

$$\dot{V} \leq \sum_{i=1}^{r} h_i [(\dot{x}^T P_i x + *) + x^T \Phi M x]$$

$$= \sum_{i=1}^{r}\sum_{j=1}^{r}\sum_{l=1}^{r}\sum_{e=1}^{r} h_i h_j h_l h_e \left[\left\{\left((A_i + \beta B_i k_j C_{2l} cos\theta_j + \beta N_i sin\theta_j)x + E_i w\right)^T P_e x \right.\right.$$
$$\left.\left. + *\right\} + x^T \Phi M x\right]$$

$$= \sum_{i=1}^{r}\sum_{j=1}^{r}\sum_{l=1}^{r}\sum_{e=1}^{r} h_i h_j h_l h_e \left[x^T\left(A_i^T + \beta C_{2l}^T k_j^T B_i^T cos\theta_j + \beta N_i^T sin\theta_j\right) P_e x \right.$$
$$+ w^T E_i^T P_e x + x^T P_e (A_i + \beta B_i k_j C_{2l} cos\theta_j + \beta N_i sin\theta_j)x \qquad (20)$$
$$\left. + x^T P_e E_i w + x^T \Phi M x\right]$$

By considering the $H_\infty$ performance level, one has:

$$\dot{V}(x) + z^T z - \gamma^2 w^T w =$$
$$\sum_{i=1}^{r}\sum_{j=1}^{r}\sum_{l=1}^{r}\sum_{e=1}^{r} h_i h_j h_l h_e \left[x^T\left(A_i^T + \beta C_{2l}^T k_j^T B_i^T cos\theta_j + \beta N_i^T sin\theta_j\right) P_e x \right.$$
$$+ w^T E_i^T P_e x + x^T P_e (A_i + \beta B_i k_j C_{2l} cos\theta_j + \beta N_i sin\theta_j)x \qquad (21)$$
$$+ x^T P_e E_i w + x^T \Phi M x + (C_{1i} x + D_i w)^T (C_{1i} x + D_i w)$$
$$\left. - \gamma^2 w^T w\right]$$

$$= \sum_{i=1}^{r}\sum_{j=1}^{r}\sum_{l=1}^{r}\sum_{e=1}^{r} h_i h_j h_l h_e \left[x^T\left(A_i^T P_e + \beta C_{2l}^T k_j^T B_i^T P_e cos\theta_j + \beta N_i^T P_e sin\theta_j \right.\right.$$
$$+ P_e A_i + \beta P_e B_i k_j C_{2l} cos\theta_j + \beta P_e N_i sin\theta_j + \Phi M)x \qquad (22)$$
$$+ x^T C_{1i}^T C_{1i} x + x^T C_{1i}^T D_i w + w^T D_i^T C_{1i} x + w^T D_i^T D_i w$$
$$\left. + w^T E_i^T P_e x + x^T P_e E_i w - \gamma^2 w^T w\right]$$

By considering Lemma 1, (22) can be rewritten as:

$$= \sum_{i=1}^{r}\sum_{j=1}^{r}\sum_{l=1}^{r}\sum_{e=1}^{r} h_i h_j h_l h_e \left[x^T\left((A_i^T P_e + *) + \varepsilon_{ijl}(C_{2l}^T k_j^T B_i^T B_i k_j C_{2l} + N_i^T N_i)\right.\right.$$
$$+ \varepsilon_{ijl}^{-1}(\beta^2 P_e^2 cos^2\theta_j + \beta^2 P_e^2 sin^2\theta_j) + C_{1i}^T C_{1i} + \Phi M\Big) x$$
$$+ (w^T(E_i^T P_e + D_i^T C_{1i})x + *) + w^T (D_i^T D_i - \gamma^2 I) w\Big]$$

$$= \sum_{i=1}^{r}\sum_{j=1}^{r}\sum_{l=1}^{r}\sum_{e=1}^{r} h_i h_j h_l h_e \left[x^T\left((A_i^T P_e + *) + \varepsilon_{ijl}(C_{2l}^T k_j^T B_i^T B_i k_j C_{2l} + N_i^T N_i)\right.\right.$$
$$+ \varepsilon_{ijl}^{-1}(\beta^2 P_e^2) + C_{1i}^T C_{1i} + \Phi M\Big) x + (w^T(E_i^T P_e + D_i^T C_{1i})x$$
$$+ *) + w^T (D_i^T D_i - \gamma^2 I)w\Big]$$

$$= \sum_{i=1}^{r}\sum_{j=1}^{r}\sum_{l=1}^{r}\sum_{e=1}^{r} h_i h_j h_l h_e \begin{bmatrix} x(t) \\ w(t) \end{bmatrix}^T \Phi' \begin{bmatrix} x(t) \\ w(t) \end{bmatrix} \qquad (23)$$

where

$$\Phi' = \begin{bmatrix} A_i^T P_e + P_e^T A_i + \varepsilon_{ijl}(C_{2l}^T k_j^T B_i^T B_i k_j C_{2l} + N_i^T N_i) + \varepsilon_{ijl}^{-1}(\beta^2 P_e^2) + C_{1i}^T C_{1i} + \Phi M & C_{1i}^T D_i + P_e^T E_i \\ * & (D_i^T D_i - \gamma^2 I) \end{bmatrix} \quad (24)$$

If $\Phi' < 0$, then $\dot{V}(x(t)) + z(t)^T z(t) - \gamma^2 w(t)^T w(t) < 0$ for all $i, j, l, e = 1, 2, \ldots, r$.

Clearly, (24) is equivalent to

$$\Phi' = \begin{bmatrix} (A_i^T P_e + *) + \varepsilon_{ijl} N_i^T N_i + C_{1i}^T C_{1i} + \Phi M & C_{1i}^T D_i + P_e^T E_i \\ * & D_i^T D_i - \gamma^2 I \end{bmatrix}$$
$$+ \begin{bmatrix} \varepsilon_{ijl}(B_i k_j C_{2l})^T B_i k_j C_{2l} & 0 \\ 0 & 0 \end{bmatrix} + \begin{bmatrix} \varepsilon_{ijl}^{-1} \beta^2 P_e^2 & 0 \\ 0 & 0 \end{bmatrix} < 0 \quad (25)$$

where $\varepsilon_{ijl} > 0$ and $\beta > 0$.

Since the previous matrix is of the quadratic matrix inequality (QMI) form, in the following, Schur complement is employed to transform the QMI to LMI. Applying Schur complement to (25) results in:

Schur complement 1:

$$H_{ijl} = \begin{bmatrix} \begin{bmatrix} (A_i^T P_e + *) + \varepsilon_{ijl} N_i^T N_i + C_{1i}^T C_{1i} + \Phi M & C_{1i}^T D_i + P_e^T E_i \\ * & D_i^T D_i - \gamma^2 I \end{bmatrix} + \begin{bmatrix} \varepsilon_{ijl}^{-1} \beta^2 P_e^2 & 0 \\ 0 & 0 \end{bmatrix} \\ * \qquad \begin{bmatrix} (B_i k_j C_{2l})^T \\ 0 \end{bmatrix} \\ \qquad -\varepsilon_{ijl}^{-1} \end{bmatrix} \quad (26)$$

where $-\varepsilon_{ijl}^{-1} < 0$.

Schur complement 2:

$$H_{ijl} = \begin{bmatrix} (A_i^T P_e + *) + \varepsilon_{ijl} N_i^T N_i + C_{1i}^T C_{1i} + \Phi M & C_{1i}^T D_i + P_e^T E_i & (B_i k_j C_{2l})^T \\ * & D_i^T D_i - \gamma^2 I & 0 \\ * & * & -\varepsilon_{ijl}^{-1} \end{bmatrix}$$
$$+ \begin{bmatrix} \varepsilon_{ijl}^{-1} \beta^2 P_e^2 & 0 & 0 \\ 0 & 0 & 0 \\ 0 & 0 & 0 \end{bmatrix} < 0 \quad (27)$$

$$\Rightarrow E_{ijl} = \begin{bmatrix} (A_i^T P_e + *) + \varepsilon_{ijl} N_i^T N_i + C_{1i}^T C_{1i} + \Phi M & C_{1i}^T D_i + P_e^T E_i & (B_i k_j C_{2l})^T & P_e \\ * & D_i^T D_i - \gamma^2 I & 0 & 0 \\ * & * & -\varepsilon_{ijl}^{-1} I & 0 \\ * & * & * & -\varepsilon_{ijl} \beta^{-2} I \end{bmatrix} < 0 \quad (28)$$

where $-\varepsilon_{ijl} \beta^{-2} < 0$.

LMI (15) is obtained. The proof is completed.

## 4 Simulation Examples

Consider the dynamics of an isothermal continuous stirred tank reactor (CSTR) for Van de Vusse reactor, Figure 1, (Chen et al., 2011; Hamdy et al., 2014; Hamdy and Hamdan, 2015) with disturbance as follows:

$$\begin{aligned}\dot{x}_1(t) &= F_1(x_1,x_2,u) = -k_1 x_1(t) - k_3 x_1^2(t) + u(t)(C_{A0} - x_1(t)) + 0.45 w_1(t) \\ \dot{x}_2(t) &= F_2(x_1,x_2,u) = k_1 x_1(t) - k_2 x_2(t) + u(t)(-x_2(t)) + 0.5 w_2(t) \\ z(t) &= 5 x_2(t) + 0.08 w_2(t) \\ y(t) &= x_2(t) \end{aligned} \quad (29)$$

where the states $x_1 (mol/L)$ and $x_2 (mol/L)$ are the concentration of the reactant inside the reactor and the concentration of the product in the output stream of the CSTR, respectively. Output $y = x_2$ determines the grade of the final product and $z(t)$ is the controlled output of CSTR. $u(t)$ is the controlled input which is the dilution rate of $u = F/V\ (h^{-1})$, where $F$ is the input flow rate to the reactor in $L/h$ and $V$ is the constant volume of the CSTR in liter. $C_{A0}$ represents the concentration of the input-feed stream to CSTR.

$w_1(t)$ and $w_2(t)$ are disturbance input and $k_1, k_2$ and $k_3$ are the kinetic parameters. The system parameters are chosen to be $k_1 = 50 h^{-1}$, $k_2 = 100 h^{-1}$ and $k_3 = 10 L/(mol.h)$, $C_{A0} = 10 mol/L$ and $V = 1L$ (Hamdy et al., 2014).

Some equilibrium points of CSTR with respect to these parameters are given in Table 1. According to these equilibrium points, $[x_e, u_e]$, which are also chosen as the desired operation points, $[x_d, u_d]$, the T–S fuzzy bilinear model is constructed (Hamdy et al., 2014).

**Table 1** Equilibrium Points

| $x_e^T$ | $u_e$ |
|---|---|
| [2.2, 0.914] | 20.3077 |
| [4.5, 1.266] | 77.7272 |
| [7.1, 0.900] | 296.2414 |

According to the method proposed in Hamdy et al. (2014) and with respect to equilibrium points $[x_e, u_e]$, matrices $A_1, A_2$ and $A_3$ and T–S fuzzy rules and robust fuzzy output feedback controller for the bilinear model will be as follows:

Rule $i$: if $x_1$ is about $M_{1i}$ then

$$\begin{cases} \dot{x}_\delta(t) = A_i x_\delta(t) + B_i u_\delta(t) + N_i x_\delta(t) u(t) + E_i w(t) \\ z_\delta(t) = C_{1i} x_\delta(t) + D_i w(t) \\ y_\delta(t) = C_{2i} x_\delta(t) \end{cases} \quad (30)$$

$$u_\delta(t) = \frac{\beta k_i y_\delta(t)}{\sqrt{1 + (k_i y_\delta(t))^2}}, \quad i = 1, 2, 3$$

where based on the method proposed in Hamdy et al. (2014) and Hamdy and Hamdan (2015), the system matrices are constructed as follows:

$$A_1 = \begin{bmatrix} -75.2383 & 7.7946 \\ 50 & -100 \end{bmatrix}, A_2 = \begin{bmatrix} -98.3005 & 11.7315 \\ 50 & -100 \end{bmatrix}$$

$$A_3 = \begin{bmatrix} -122.1228 & 8.8577 \\ 50 & -100 \end{bmatrix}$$

$$B_1 = B_2 = B_3 = \begin{bmatrix} 10 \\ 0 \end{bmatrix}$$

$$N_1 = N_2 = N_3 = \begin{bmatrix} -1 & 0 \\ 0 & -1 \end{bmatrix}$$

$$E_1 = E_2 = E_3 = \begin{bmatrix} 0.45 & 0 \\ 0.1 & 0.5 \end{bmatrix}$$

$$D_1 = D_2 = D_3 = [0 \quad 0.08]$$

$$C_{11} = C_{12} = C_{13} = [0 \quad 5]$$

$$C_{21} = C_{22} = C_{23} = [0 \quad 1]$$

and $x_\delta = x - x_d$, $z_\delta = z - z_d$, $u_\delta = u - u_d$, $y_\delta = y - y_d$, $i = 1, 2, 3$. The membership functions are defined as follows:

$$M_{11} = \exp(-(x_1 - 2.2)^2), \quad M_{12} = \exp(-(x_1 - 4.5)^2)$$

$$M_{13} = \exp(-(x_1 - 7.1)^2) \tag{31}$$

According to our proposed control scheme, the control design procedure will be as the following: First, let $\rho = 0.1$, $\gamma = 0.3$, $\varepsilon_{ijl} = 1$ in LMI (15). Then, by applying Theorem 1 and solving the related LMIs (13), (14) and (15) via the LMI toolbox, one can figure out the positive definite matrices $P_1, P_2$ and $P_3$ as follows:

$$P_1 = \begin{bmatrix} 27.9685 & -7.0275 \\ -7.0275 & 18.5635 \end{bmatrix}, \quad P_2 = \begin{bmatrix} 27.9685 & -7.0275 \\ -7.0275 & 18.5635 \end{bmatrix}$$

$$P_3 = \begin{bmatrix} 27.9685 & -7.0275 \\ -7.0275 & 18.5635 \end{bmatrix}$$

and the controller gains as:

$$K_1 = [-2.2288], \quad K_2 = [-2.2288], \quad K_3 = [-2.2288].$$

Now by considering data from the previous steps, one can construct the fuzzy controller (5). By applying the fuzzy output feedback controller (5) to system (7), under initial condition $x(0) = [3.1 \quad 1.5]^T$ and the exogenous disturbance input $w(t) = [2e^{-0.001t}\sin(3t) \quad 3e^{-0.001t}\sin(0.1t)]^T$, we can obtain the following results.

Figures 2 and 3 show the simulation results of applying the robust output feedback controller (30) to the fuzzy bilinear system (FBS) for zero equilibrium state and for desired equilibrium state $[4.5 \quad 1.266]^T$, respectively. As these figures illustrate, the system states converge to the equilibrium in less than 0.1 seconds and the maximum absolute value of control input is less than 0.1 which shows the good performance of designed controller.

The simulation results of Figure 4 show the state trajectories for zero equilibrium with respect to three different initial states, where all states converges to equilibrium states very fast after about 0.08 sec.

The robust behavior of the system is investigated in Figure 5, where the state trajectories are changed from zero equilibrium state to $x_d = [4.5 \quad 1.266]^T$ at $t = 0.4\ sec$ three different initial conditions $x(0) = [3.1 \quad 1.5]^T$ solid line, $x(0) = [0.5 \quad 0.6]^T$ dash-dotted line and $x(0) = [-1.2 \quad -3.1]^T$ dashed line, respectively. One can find from these figures that the proposed robust output feedback controller has an excellent tracking performance and stability of the system is completely satisfactory since the states converges to the equilibrium states in less than 0.1 second.

To show the advantage of our work better, some comparative studies is done. The simulation results of applying the robust fuzzy controller (5) to the original nonlinear system and FBS (30) under initial condition $x(0) = [3.1\ 1.5]^T$ are shown in Figure 6. One can see from this figure that the state responses of the original nonlinear system and the proposed fuzzy output feedback controller under the same control input are almost the same without any overshoot. By considering results from Figure 6 and that of references Li and Tsai (2007), Tsai (2011) and Tsai et al. (2015) which are based on state feedback control, one can find that our proposed approach shows less overshoot and oscillation than those of proposed methods in references Li and Tsai (2007), Tsai (2011) and Tsai et al. (2015).

## 5 Conclusions

A new robust $H_\infty$ fuzzy control scheme for a class of bilinear system with disturbance based on NQLF approach via PDC technique has been proposed in this paper. Moreover, by considering the upper bounds of MFs as an LMI variable and utilizing a new slack matrix, the stability conditions have been obtained in LMIs. The proposed robust fuzzy output feedback controller can guarantee a prescribed level on disturbance attenuation. Finally, some examples have been utilized to demonstrate the effectiveness of the proposed fuzzy model and controller via CSTR benchmark. As the future work, we can apply the proposed model to different benchmarks and compare the results with the other methods.

## Figures

**Figure 1**  Diagram of the CSTR reactor

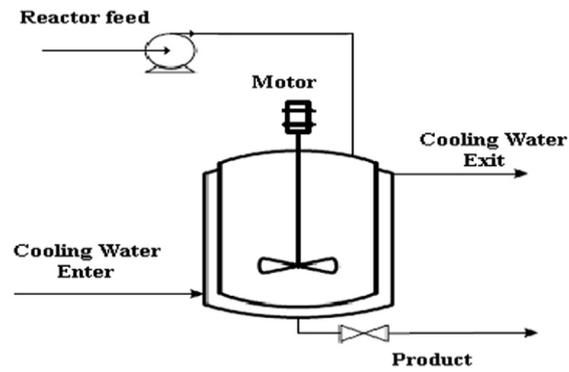

**Figure 2**  State responses of **(a)** FBS and **(b)** control trajectory of system under zero equilibrium state

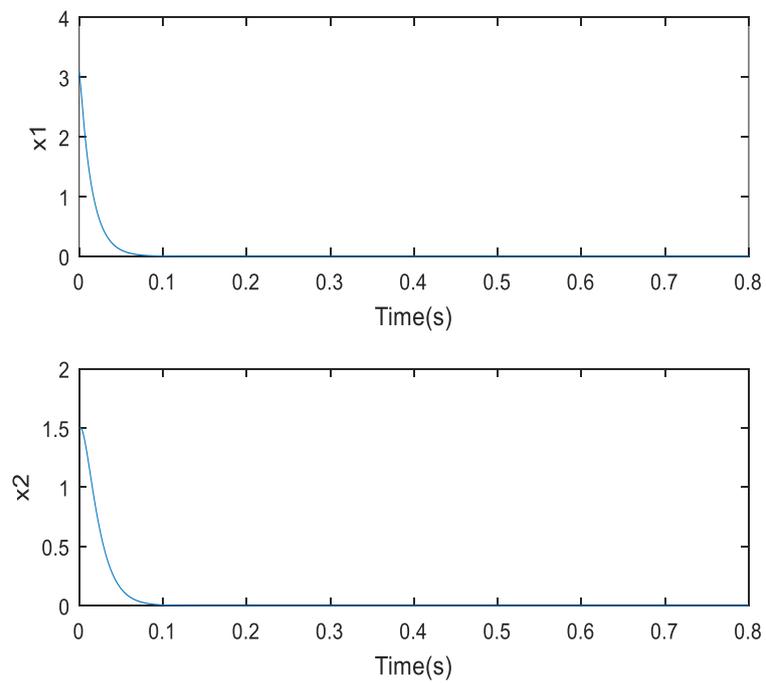

(a)

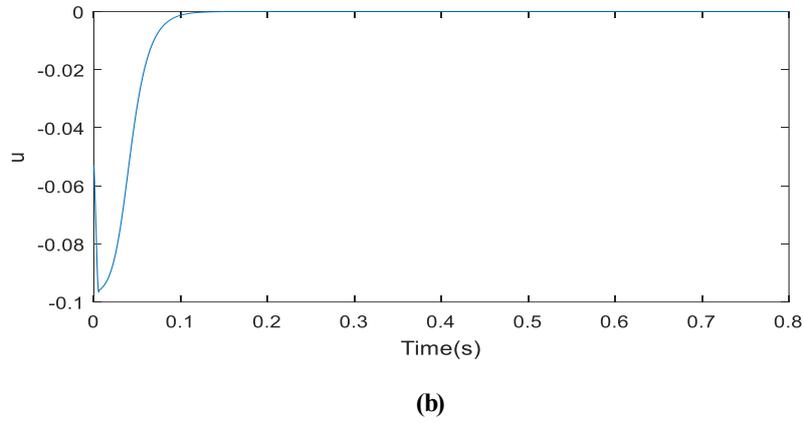

(b)

**Figure 3** State responses of FBS for equilibrium state of $[4.5 \ \ 1.266]^T$

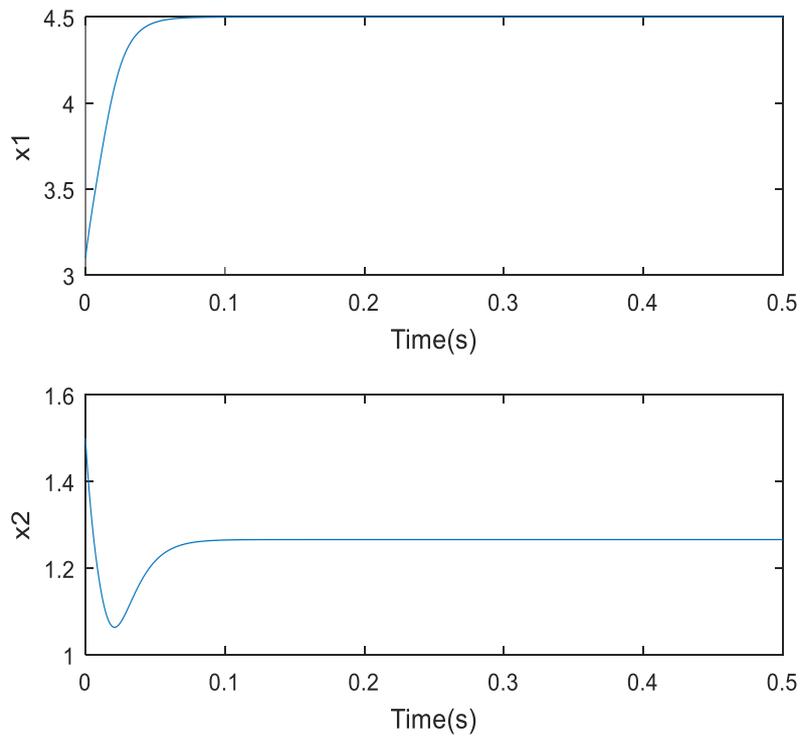

**Figure 4** State responses of FBS under tree different initial conditions, $x(0) = [3.1 \ 1.5]^T$ solid

line, $x(0) = [0.5 \ 0.6]^T$ dash-dotted line and $x(0) = [-1.2 \ -3.1]^T$ dashed line.

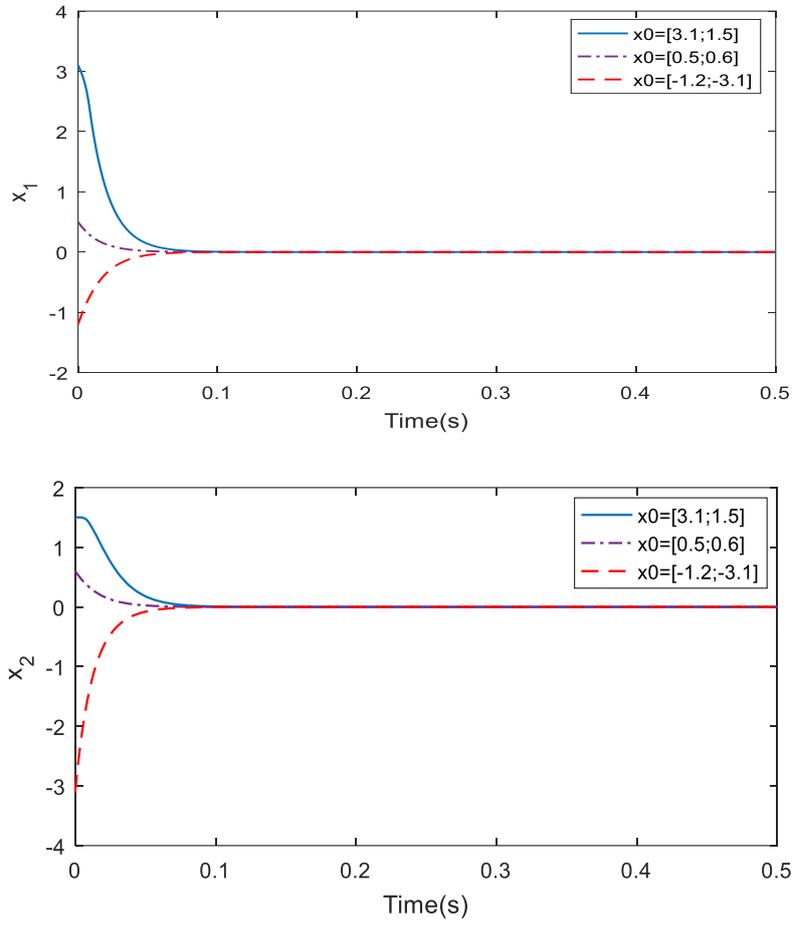

**Figure 5** State responses of FBS for zero equilibrium state and $[4.5 \ 1.266]^T$ equilibrium state under different initial conditions

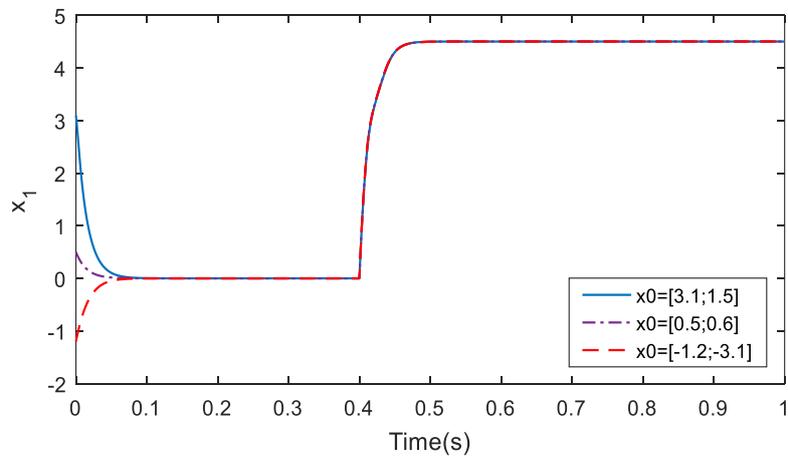

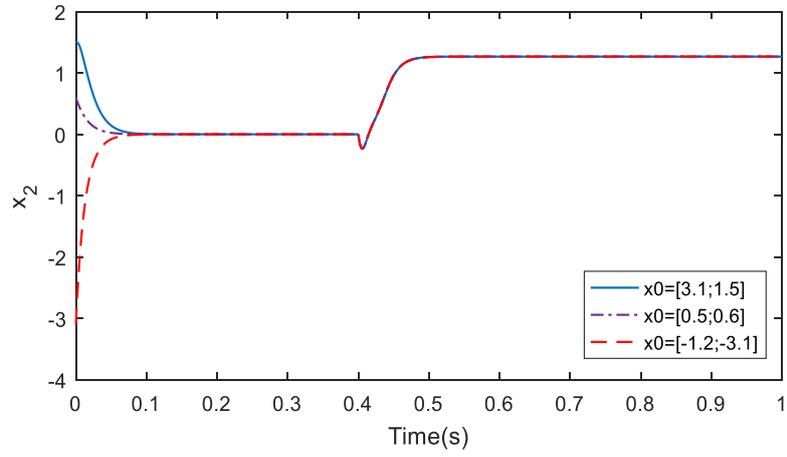

**Figure 6** State responses the nonlinear model of CSTR, dashed line and the proposed FBS, solid line

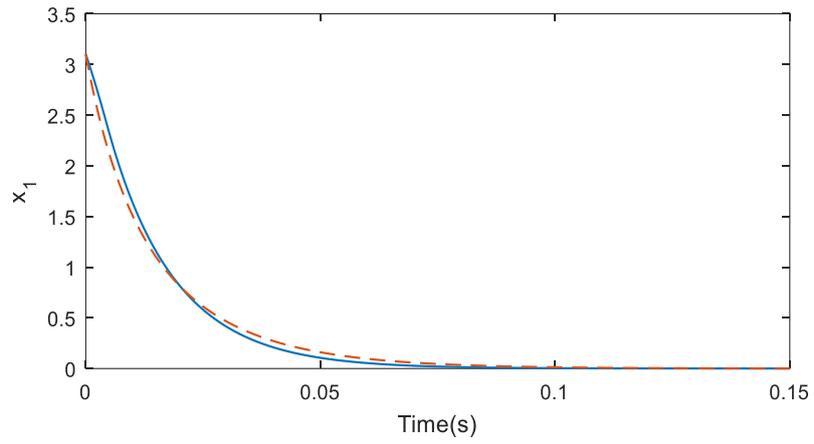

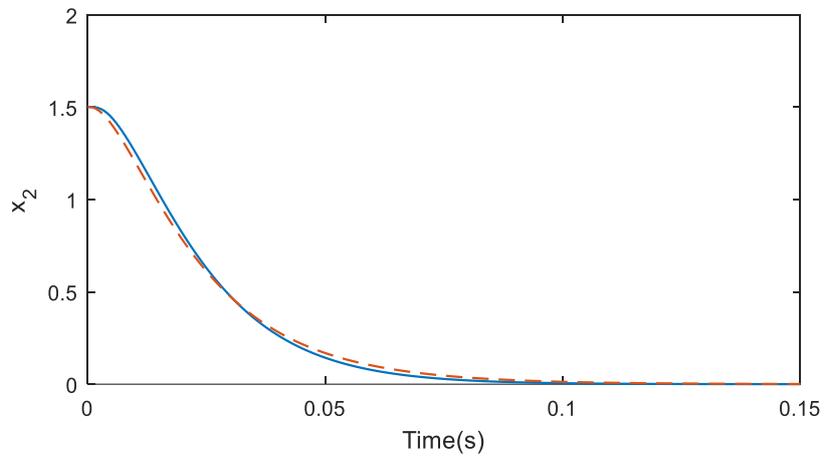